\documentclass[11pt]{article}
\usepackage{moriond}
\usepackage{lineno}

\def\Journal#1#2#3#4{{#1} {\bf #2}, #3 (#4)}

\def\NPB{{\em Nucl. Phys.} B}
\def\PLB{{\em Phys. Lett.}  B}
\def\PRL{\em Phys. Rev. Lett.}
\def\PRD{{\em Phys. Rev.} D}
\def\EPJC{{\em Eur. Phys. J.} C}
\def\JHEP{\em JHEP}
\def\JINST{\em JINST}

\def\be{\begin{equation}}
\def\ee{\end{equation}}
\def\bea{\begin{eqnarray}}
\def\eea{\end{eqnarray}}

\newcommand{\Photo}{}

\begin{document}
\vspace*{4cm}
\title{HEAVY FLAVOR MEASUREMENTS AT LHC \footnote{Prepared for the proceedings of Les Rencontres de Moriond 2013, Electroweak Session.}}

\author{S. Spagnolo$^\star$ on behalf of the ATLAS and CMS Collaborations}

\address{$^\star$ INFN Lecce and Dipartimento di Matematica e Fisica ``Ennio De Giorgi'', Universit\`a del Salento, \\
Lecce, Italy}

\maketitle\abstracts{
The measurements in the area of heavy flavor physics, produced by Winter 2013 by the ATLAS and CMS experiments at LHC, are reviewed with focus on the most recent results. 
}

\section{Introduction}
The majority of the results on heavy flavor physics produced so far by the ATLAS\cite{bib:ATL} and CMS\cite{bib:CMS} experiments are based on the limited statistics of about $\rm 40~pb^{-1}$ per experiment collected in the 2010 LHC run at $\rm \sqrt{s}=7~TeV$. The relatively low luminosity and pileup allowed to select interesting events with inclusive low transverse momentum single or di-muon triggers, exploiting semileptonic decays of heavy flavors and $J/\psi\rightarrow\mu^+\mu^-$ decays. The measurements based on the 2011 data set, consisting of about $\rm 5~fb^{-1}$ per experiment, with an average number of interactions per crossing ranging from 6 to 12, have been performed thanks to dedicated $J/\psi$, $\Upsilon$ and $B\rightarrow\mu^+\mu^-$ triggers, with invariant mass selection and common vertex fit for the two muons. The main topics that have been investigated so far are inclusive and exclusive heavy flavor production, properties of well known and new hadrons with beauty, studies of the $c\bar{c}$ and $b\bar{b}$ bound states, precision measurements of the $B_s$ system for CP violation studies and searches for rare decays. The results, summarized in this report, anticipate more and better contributions to the understanding of heavy flavor physics which will come with the analysis of the larger and still unexploited 2012 data set.

\section{Inclusive heavy flavor production}
\label{sec:incl}
Several measurements of inclusive heavy flavor production were produced by ATLAS and CMS: inclusive differential cross section of muons ($\mu$) and electrons ($e$) from $c$- and $b$-quark decays\cite{bib:ATLinclusivemue}, $b$-jet inclusive production via secondary vertex identification or semileptonic decays to muons\cite{bib:CMSinclusivebjet,bib:ATLinclusivebjet} and $b$-jet inclusive pair-production\cite{bib:ATLinclusivebjet,bib:CMSinclusivebbsigma}. 
\begin{figure}[t]
\begin{minipage}{0.28\linewidth}
\centerline{\includegraphics[width=0.95\linewidth]{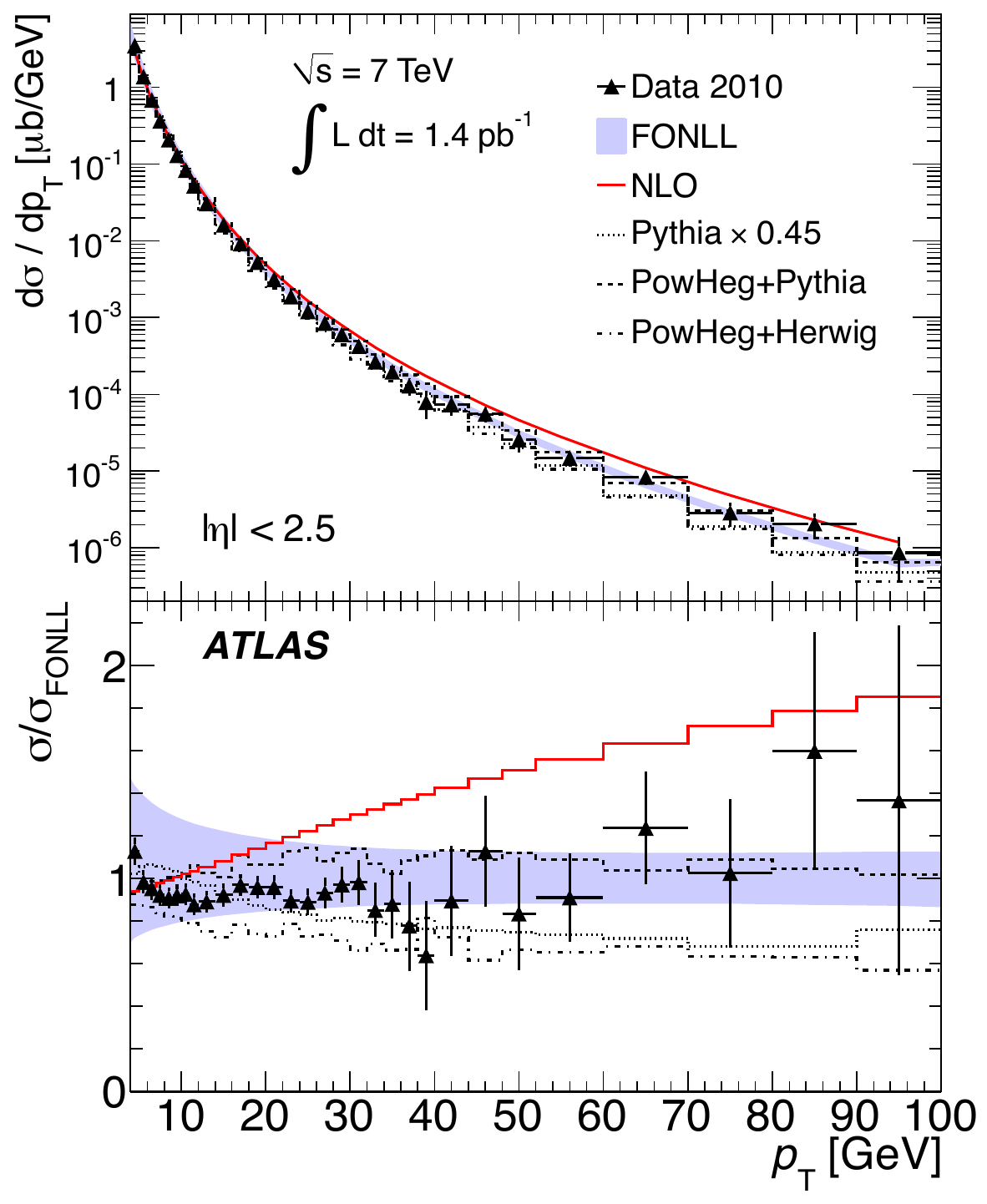}}
\end{minipage}
\begin{minipage}{0.3\linewidth}
\centerline{\includegraphics[width=1.0\linewidth]{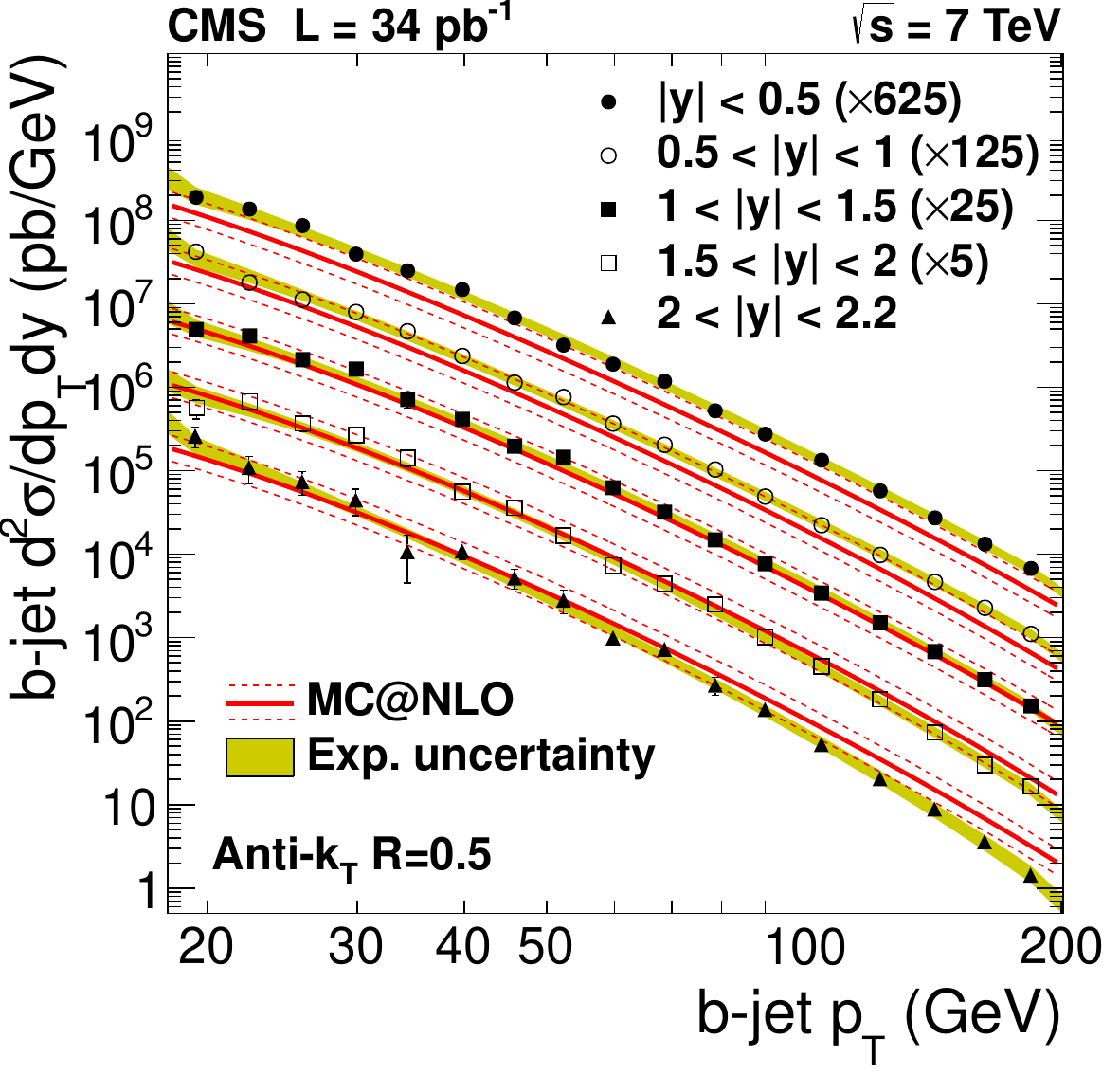}}
\end{minipage}
\begin{minipage}{0.4\linewidth}
\centerline{\includegraphics[width=1.0\linewidth]{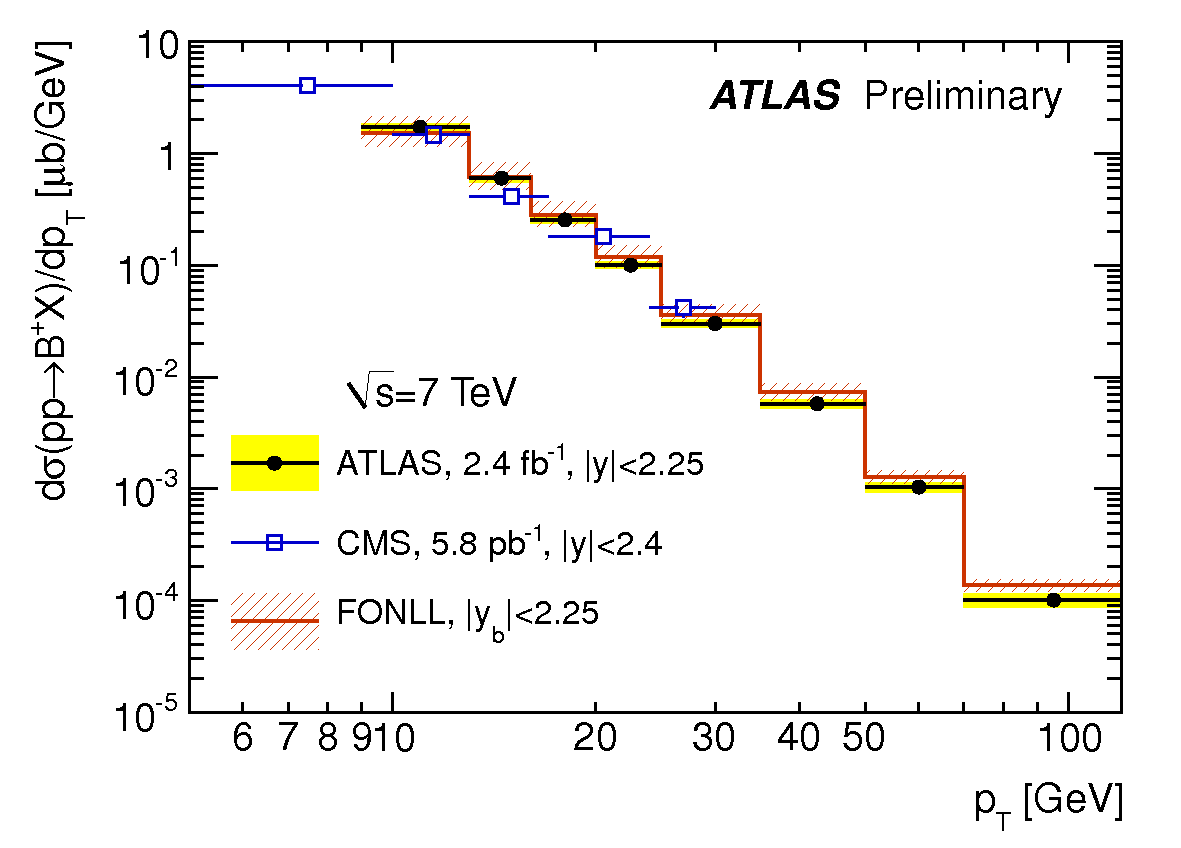}}
\end{minipage}
\vspace{-0.2cm}
\caption{
Left: ATLAS inclusive differential production cross section for muons from heavy flavor decays\protect\cite{bib:ATLinclusivemue}. Center: CMS inclusive $b$-jet production cross section as a function of $\rm p_T$ for several rapidity bins\protect\cite{bib:CMSinclusivebjet}. Right: differential $B^\pm$ production cross section measured by ATLAS\protect\cite{bib:ATLBplusXsec}, compared with theory predictions and with CMS results.}
\label{fig:1}
\end{figure}
They have shown that QCD NLO matrix element calculation and next to leading logarithms resummation at high $\rm p_T$ (as implemented, for example, in the {\tt FONNL} calculation\cite{bib:FONNL}) are necessary in order to describe the data with good accuracy. 
Inclusive $b$-pair production is well reproduced by all Monte Carlo (MC) generators. The double differential $\rm p_T,y$ distribution of inclusive $b$-jets is well predicted by {\tt POWHEG}\cite{bib:powheg},  
based on a NLO matrix element matched to the {\tt PYTHIA}\cite{bib:pythia} parton shower.  
On the other hand, 
{\tt MC$@$NLO}\cite{bib:MCatNLO}, implementing the same fixed order pQCD approximation but interfaced to the {\tt HERWIG}\cite{bib:herwig} parton shower, fails to predict the inclusive $b$-jet rate in some regions of the phase space. 
A few distributions from these studies are shown in Fig.\ref{fig:1} (left and center). 
The detailed modeling of events with heavy quarks, which are often important backgrounds for searches of new phenomena and Higgs studies, requires in the generators the tuning of phenomena not described by perturbative QCD. A variety of measurements\cite{bib:ATLDstarinjets} hint to mis-modeling of specific aspects, like $b$-fragmentation and relative relevance of heavy flavor production mechanisms, and they offer a valuable input for MC tuning. 

\section{Exclusive B-hadron measurements and rare states}
\label{sec:excl}
The production cross section for $B^+, B^0$ and $B_s$ was measured\cite{bib:CMSexclBsigma} with the 2010 data by CMS and compared with the {\tt MC$@$NLO} predictions, which are typically underestimating the measured values, although in agreement with them within the theory uncertainties (currently larger than the experimental errors). A recent high statistics measurement of the differential production cross section of $B^+$ has been produced with $\rm 2.4~fb^{-1}$ of data collected in 2011 by ATLAS\cite{bib:ATLBplusXsec}. A large sample of $B^\pm$ has been reconstructed through the decay in $J/\psi(\mu^+\mu^-)K^\pm$; fiducial measurements of  $\rm d\sigma/dp_T$ (see Fig.\ref{fig:1}, right) and $\rm d\sigma/d|y|$ have been obtained after correcting the event yield for detector efficiency and acceptance. The comparison with theory predictions confirms the good agreement between data and the {\tt FONNL} calculation as well as the {\tt POWHEG} generator interfaced with {\tt PYTHIA}.
\begin{figure}[b]
\begin{minipage}{0.3\linewidth}
\centerline{\includegraphics[width=.9\linewidth]{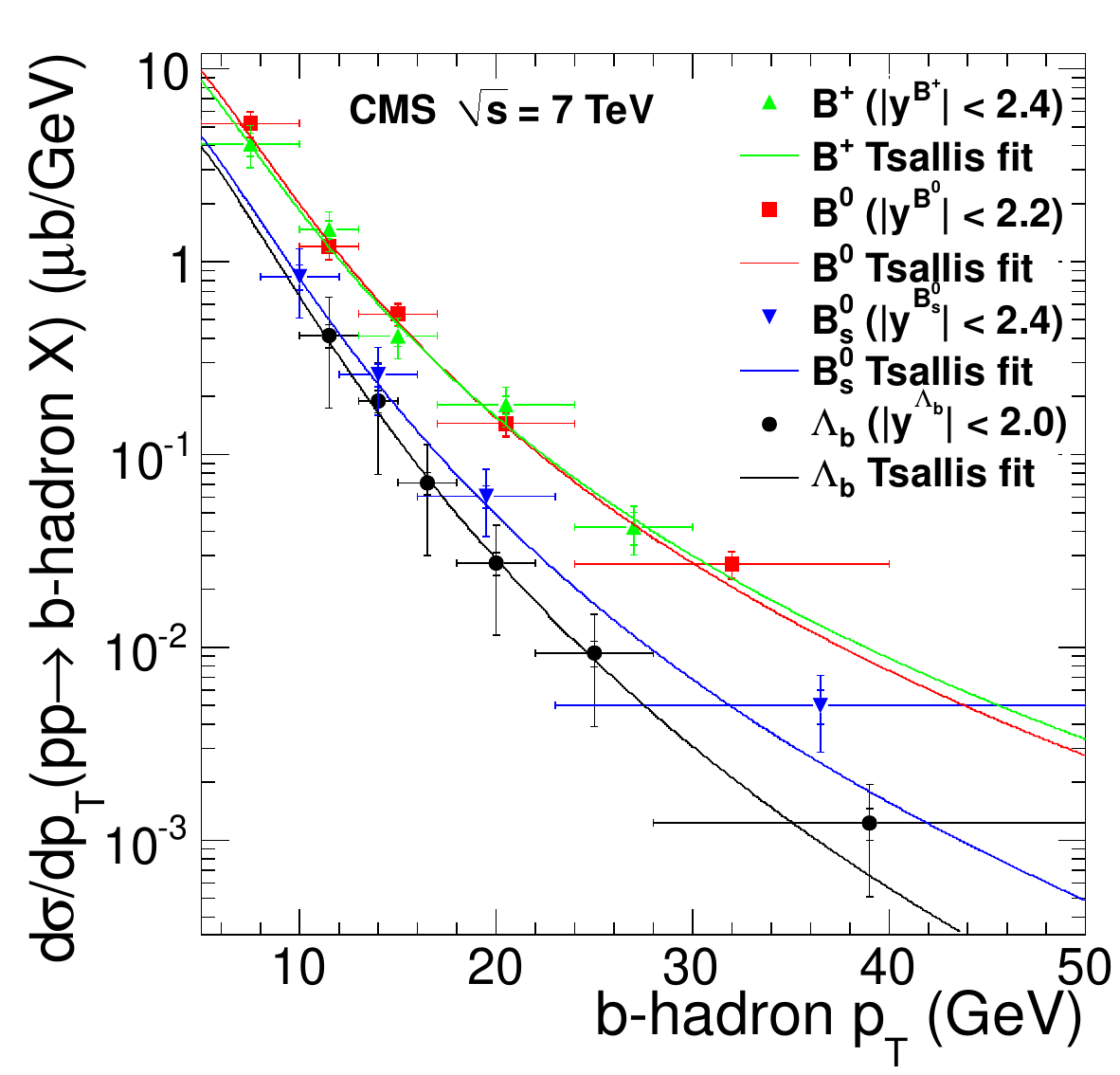}}
\end{minipage}
\begin{minipage}{0.38\linewidth}
\centerline{\includegraphics[width=1.0\linewidth]{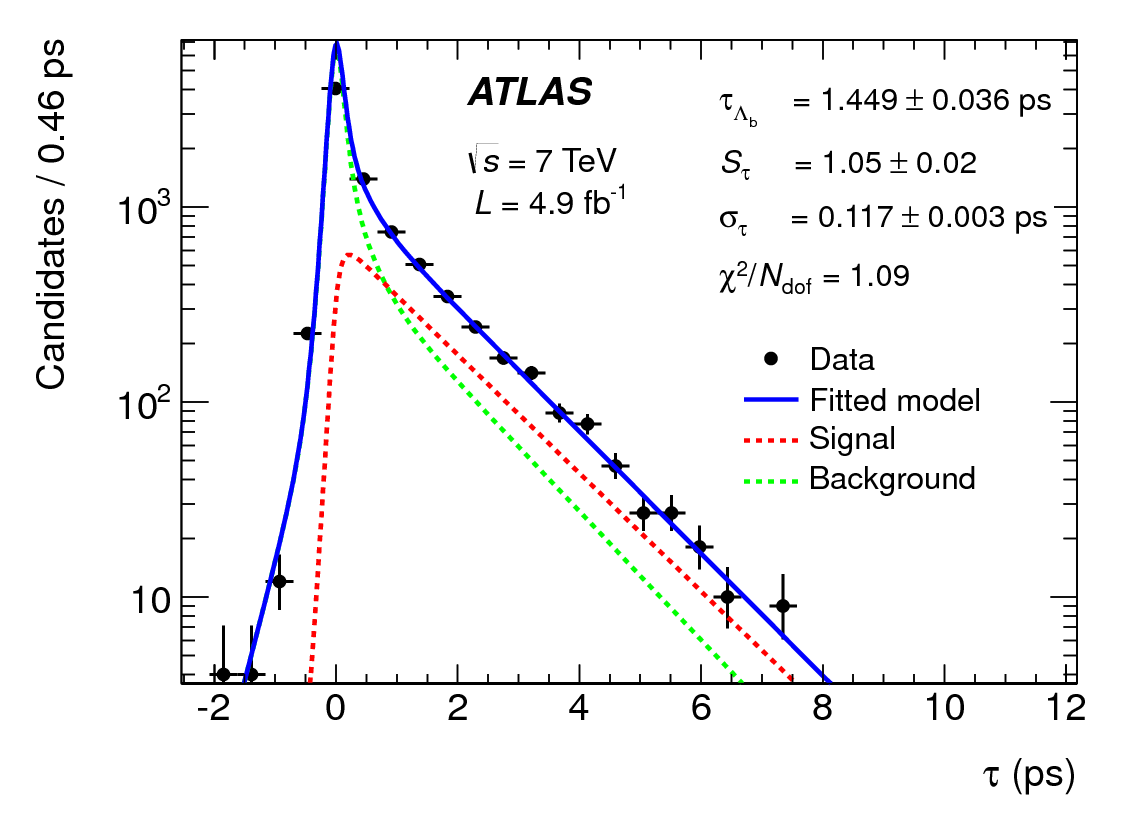}}
\end{minipage}
\begin{minipage}{0.3\linewidth}
\centerline{\includegraphics[width=0.95\linewidth]{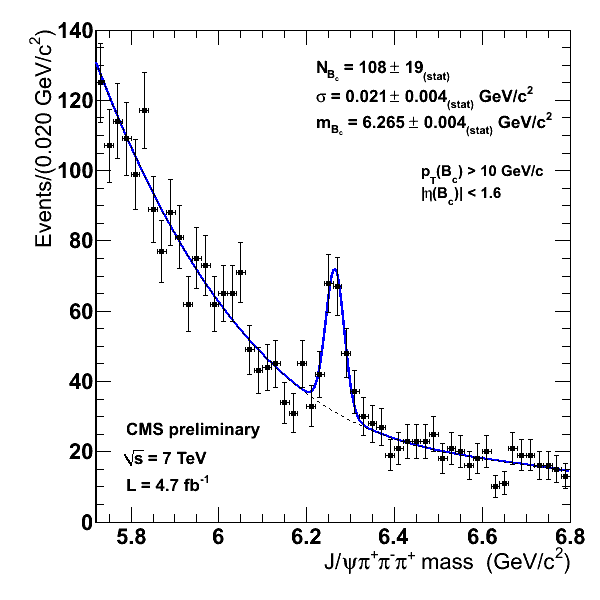}}
\end{minipage}
\vspace{-0.2cm}
\caption{Left: CMS measurement\protect\cite{bib:CMSlambdabsigma} of $\rm d\sigma/dp_T$ for $\Lambda_b$, exhibiting a steeper $\rm p_T$ dependence than observed for B-mesons. Center: ATLAS projection of the fit results on the proper decay time distribution of $\Lambda_b$ candidates selected for the measurement of the lifetime\protect\cite{bib:ATLLambdabmt}. Right: CMS observation\protect\cite{bib:CMSbc} of $\rm B_c$ in the $J\psi\pi^+\pi^-\pi^+$ decay.}
\label{fig:2}
\end{figure}

The large sample of 2011 has been exploited by the LHC experiments for measurements of rarer $b$-hadrons, like $\Lambda_b$\cite{bib:CMSlambdabsigma,bib:ATLLambdabmt,bib:CMSLambdabt}. $\Lambda_b$ is reconstructed through the four particle final state $\mu^+\mu^-p\pi^-$ reached via the decay to $\Lambda J/\psi$. The main background, coming from $B^0_d\rightarrow J/\psi(\mu^+\mu^-)K_s(\pi^+\pi^-)$, which has a similar event topology, is rejected via a direct veto on the $K_s$ mass. The signal yield is extracted from a fit to the  $\Lambda_b$  peak in the invariant mass distribution emerging over the combinatorial background, dominated by $J/\psi$ from $B$ decays. CMS measured the  differential production cross section\cite{bib:CMSlambdabsigma} of $\Lambda_b$ times the branching fraction to $\Lambda J/\psi$ 
and the ratio $\sigma(\bar{\Lambda}_b)/\sigma(\Lambda_b)$ as a function of $y$. The $\rm p_T$ spectrum is found to be softer than for B mesons (see Fig.\ref{fig:2}, left); this observation suggests a dependency of the $b$-quark fragmentation fraction on $\rm p_T$ which might explain the apparent disagreement between the measurement of the fragmentation fraction to $\Lambda_b$ at LEP and at Tevatron, where the $b$-quarks have an average momentum lower than in jets from Z decays.  No baryon/anti-baryon asymmetry is observed within the statistical and systematic errors.
ATLAS performed a measurement\cite{bib:ATLLambdabmt} of the mass and lifetime of $\Lambda_b$, with an unbinned maximum likelihood fit to the candidate invariant mass and proper decay time.  In Fig.\ref{fig:2} (center) the projection of the fit result in the proper decay time distribution and the data are shown. The measurements are in agreement with the previous world average values and the lifetime from ATLAS, $\rm \tau_{\Lambda_b}=1.449\pm 0.036(stat)\pm 0.017(syst)~ps$, is currently the most precise determination of this parameter. A preliminary measurement\cite{bib:CMSLambdabt} of $\tau_{\Lambda_b}$ by CMS is consistent with the ATLAS result.  ATLAS and CMS collected evidence\cite{bib:CMSbc,bib:ATLbc} for the rare and interesting $B_c$ meson, consisting of a b and a c quark. After selecting an opposite-sign muon pair and a $\pi$ track originating from a common secondary vertex and fitting the kinematics of the three particle vertex, a peak in the  $J/\psi\pi^\pm$ invariant mass distribution is observed with a significance of 5 and 10 $\sigma$ in ATLAS and CMS respectively. In addition, CMS\cite{bib:CMSbc} has observed, with a significance of 6 $\sigma$, the signal in the decay channel $J/\psi\pi\pi\pi$ as shown in Fig.\ref{fig:2} (right).
\begin{figure}[t]
\begin{minipage}{0.3\linewidth}
\centerline{\includegraphics[width=0.9\linewidth]{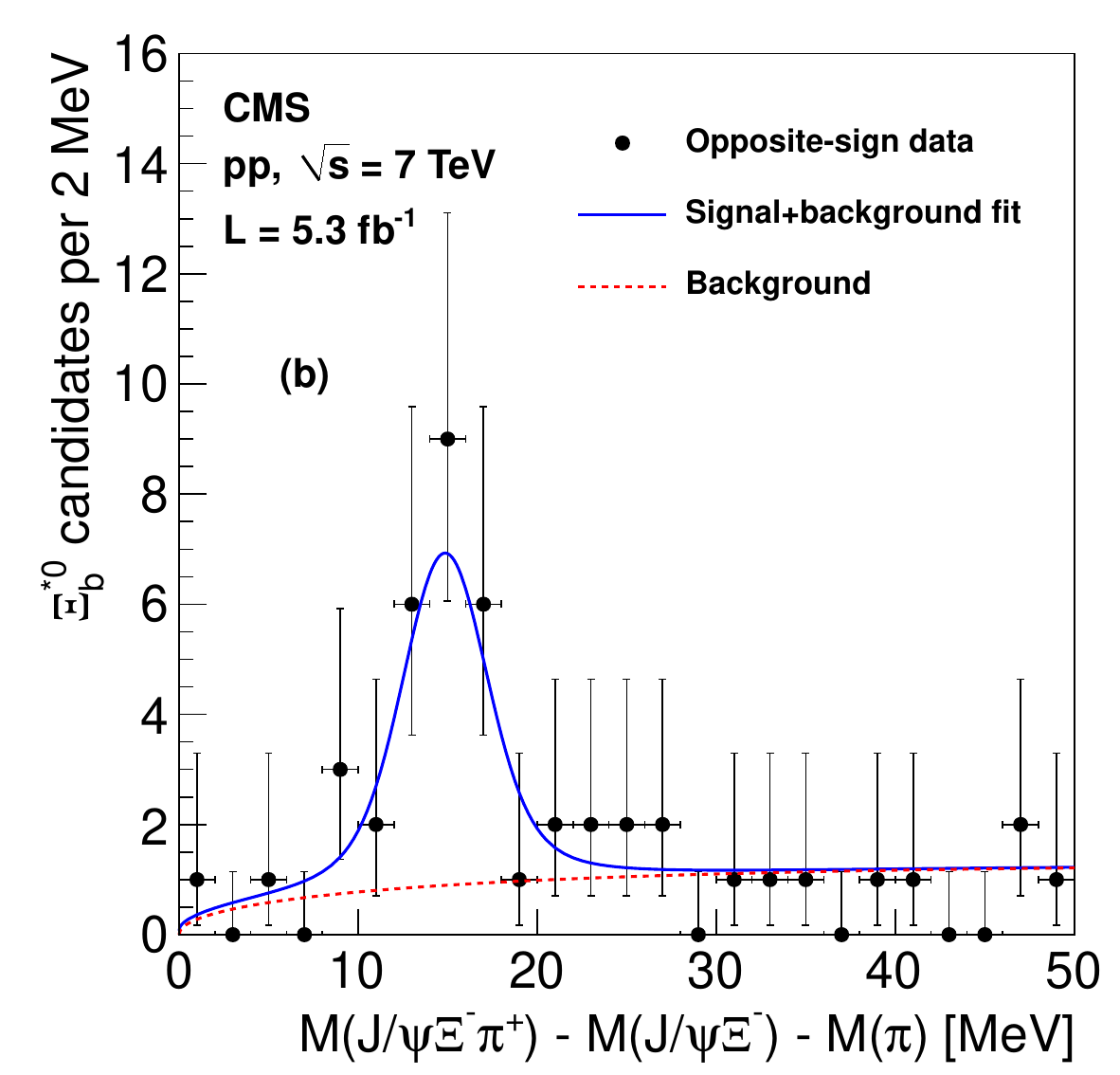}}
\end{minipage}
\begin{minipage}{0.35\linewidth}
\centerline{\includegraphics[width=0.95\linewidth]{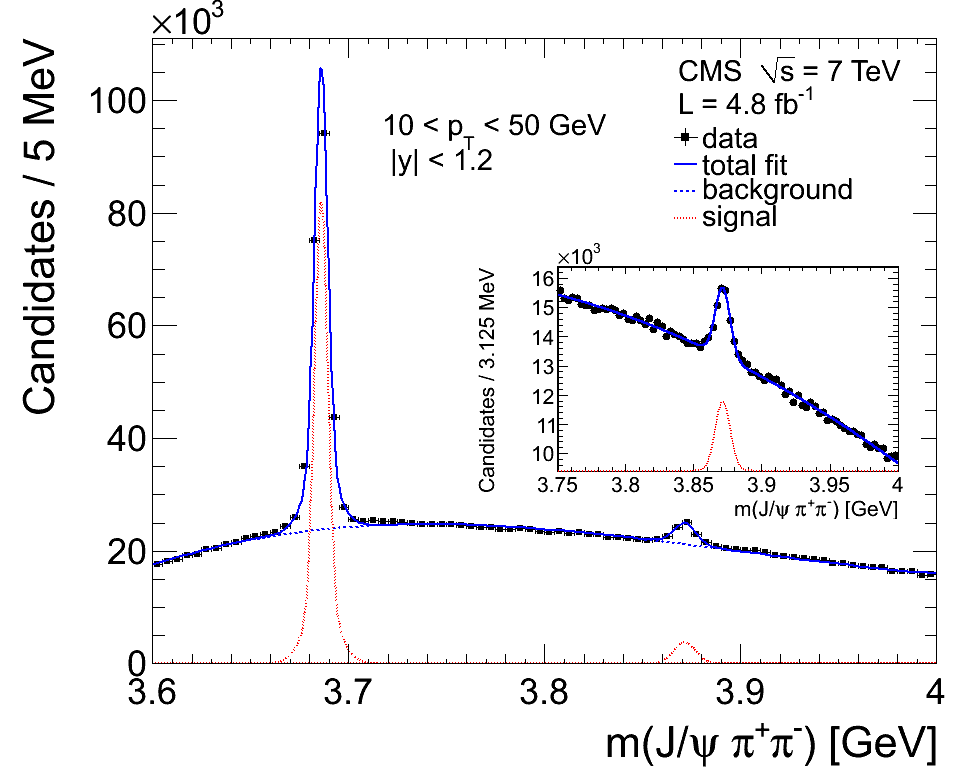}}
\end{minipage}
\begin{minipage}{0.35\linewidth}
\centerline{\includegraphics[width=1.0\linewidth]{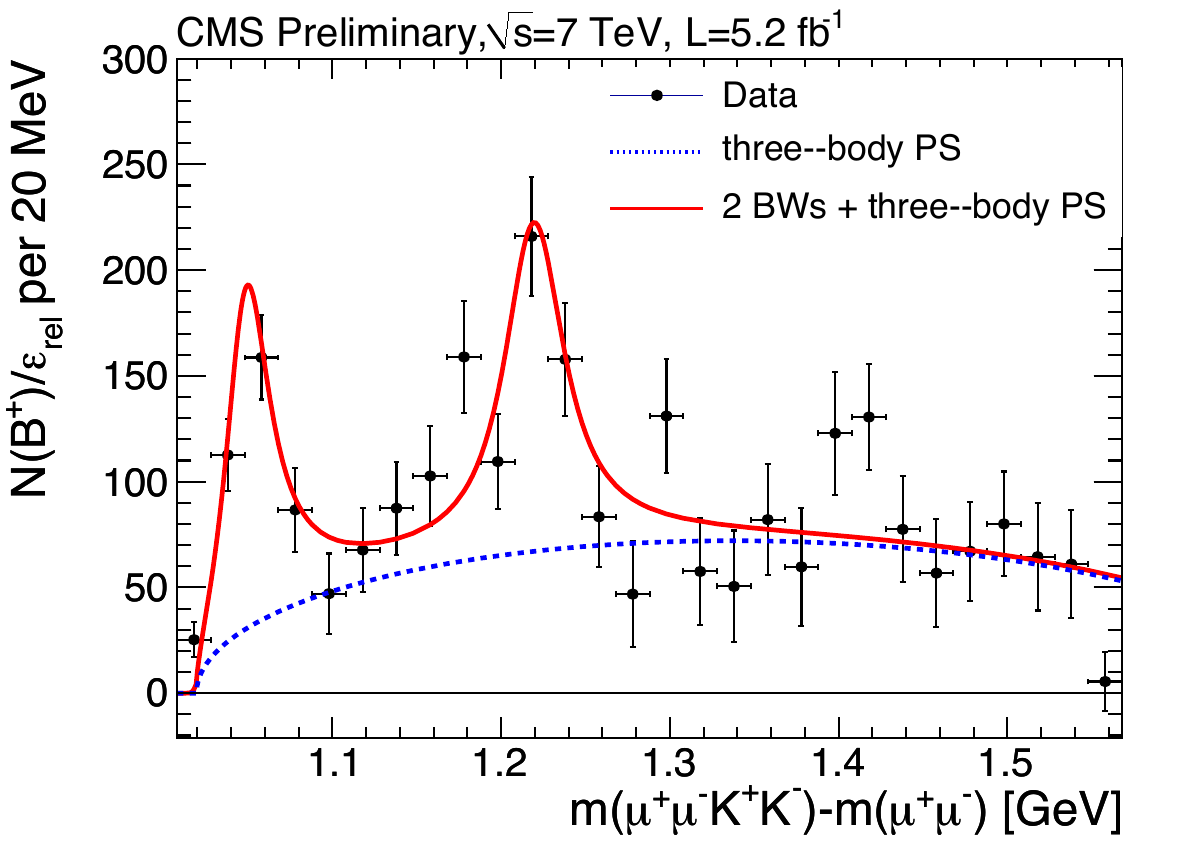}}
\end{minipage}
\caption{Left: Distribution of the Q-value for the first observation of $\Xi_b^\star\rightarrow\Xi_b(J/\psi\Xi_b)\pi$ from CMS\protect\cite{bib:CMSxib}. Center: CMS $\rm X(3872)$ signal in $J/\psi\pi^+\pi^-$ used to measure the differential production cross section\protect\cite{bib:CMSX3872} of $\rm X(3872)$ relative to $\psi(2S)$. Right: CMS evidence\protect\cite{bib:CMSstruct} for structures in the $J/\psi\phi$ spectrum from $B^+\rightarrow J/\psi\phi K^+$ decays.}
\label{fig:3}
\end{figure}

A first observation of a new baryon of the $\Xi_b$ family was performed by CMS. A decay chain with three dispalced decay vertices, involving charged particles only, has been used to isolate the signal corresponding to the strong decay to  $\Xi_b^-\pi^+$ of the $\Xi_b^\star$ candidate state, foreseen by the quark model with quantum numbers $\rm J^{P}=3/2^+$. The $\Xi_b^-\rightarrow\Xi^-(\Lambda\pi^-) J/\psi$ decay, with $\Lambda\rightarrow\pi p$ and $J/\psi\rightarrow\mu\mu$ is fully reconstructed with constraints on the mass of the decaying particles at each vertex; the distribution of the variable $Q= M(J/\psi \Xi \pi^+)-M(J/\psi \Xi)-M(\pi^+)$, reported in Fig.\ref{fig:3} (left), is well fitted with a Breit-Wigner with free mass and width which are determined with values consistent with the predictions of Lattice QCD. In addition, the state of unclear interpretation X(3872) has been studied through the decay in $J/\psi\pi^+\pi^-$ with the 2011 CMS data\cite{bib:CMSX3872}. The production rate as a function of $\rm p_T$, normalized to the yield of $\psi(2S)$ in the same decay channel, and the ratio of non prompt to prompt production have been measured. These measurements will hopefully help to clarify the nature of this particle which doesn't easily fit into the quark model. 
Finally, CMS has reported evidence for resonant  structures in the $J/\psi \phi$ spectrum, when looking at decays of $B^\pm\rightarrow J/\psi\phi K^\pm$. The distribution of $\rm Q=M(\mu^+\mu^- K^+K^-)-M(\mu^+\mu^-)$, reported in Fig.\ref{fig:3} (right) can be fitted with two Breit-Wigner, superimposed to the smooth 3-body phase-space, at mass values of $\rm 4148.2\pm 2.0(stat)\pm 4.6(syst)~MeV$ and $\rm 4316.7\pm 3.0(stat)\pm 7.3(syst)~MeV$. The first peak, which has a significance exceeding 5 $\sigma$, is consistent with a preliminary claim by CDF not yet  confirmed.

\section{Onia properties}
\label{sec:onia}
\begin{figure}[t]
\begin{minipage}{0.3\linewidth}
\centerline{\includegraphics[width=1.0\linewidth]{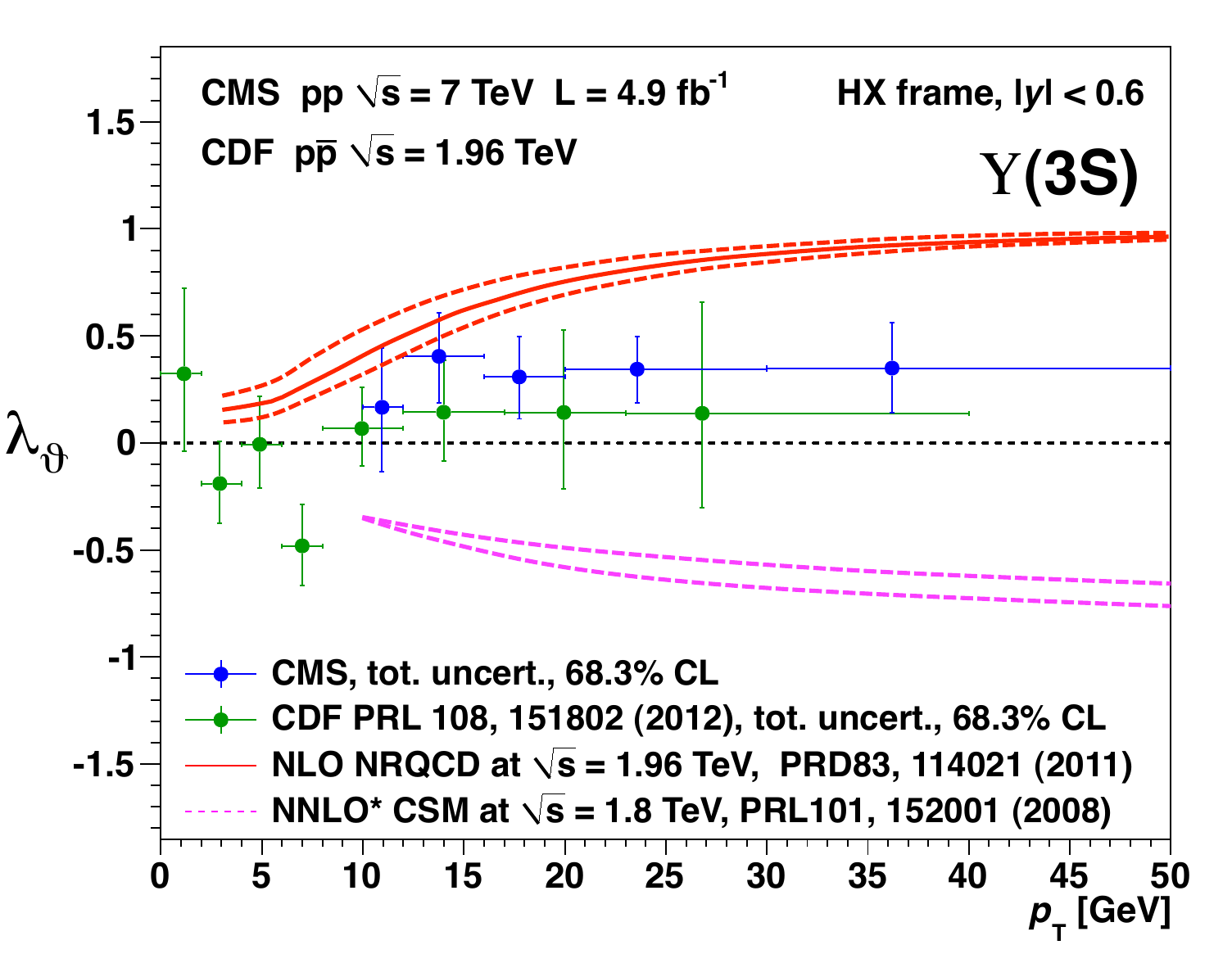}}
\end{minipage}
\begin{minipage}{0.38\linewidth}
\centerline{\includegraphics[width=0.95\linewidth]{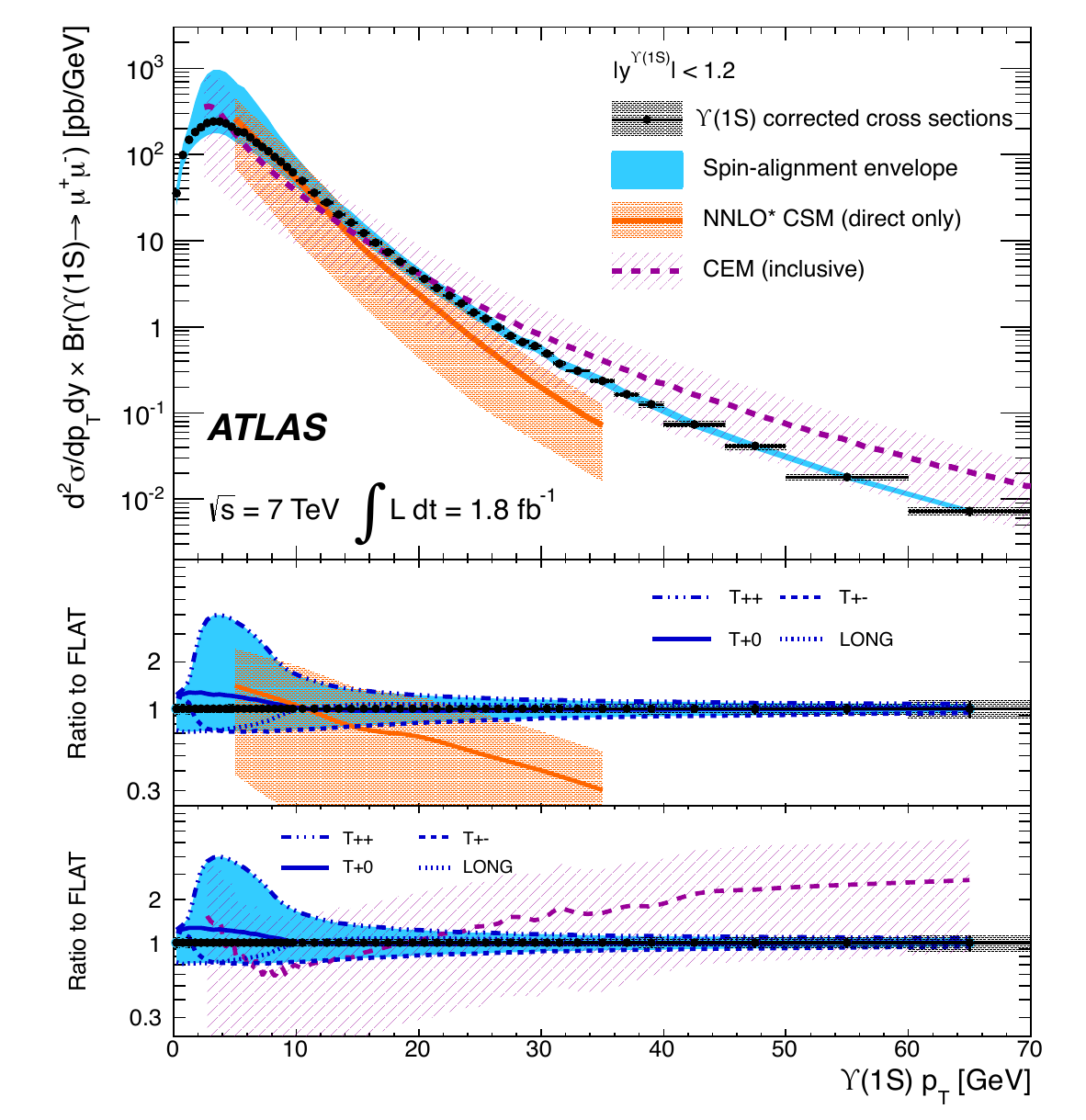}}
\end{minipage}
\begin{minipage}{0.3\linewidth}
\centerline{\includegraphics[width=1.0\linewidth]{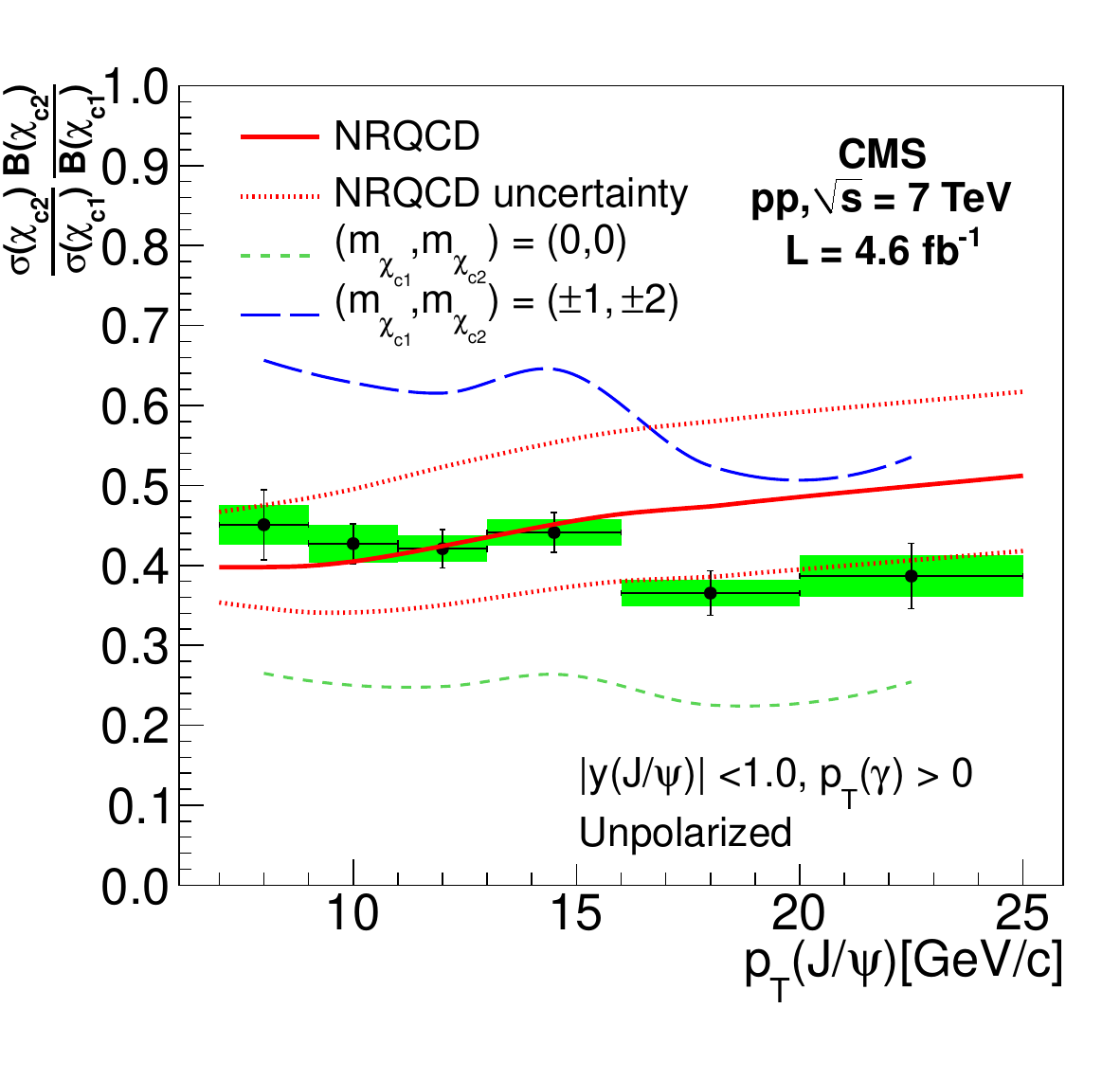}}
\end{minipage}
\vspace{0.cm}
\caption{Left: CMS measurement\protect\cite{bib:CMSYpol} of the $\lambda_\theta$ polarization parameter for $\Upsilon(3S)$ superimposed to previous CDF measurements and various NRQCD predictions. Center: ATLAS $\rm p_T$ differential production cross secton for $\Upsilon(1S)$ produced at central rapidity\protect\cite{bib:ATLYxsec}; perturbative QCD predictions and expectations from phenomenological models are superimposed to the data. 
 Right: ratio of $\chi_{c2}$ and $\chi_{c1}$ production cross section measured by CMS\protect\cite{bib:CMSchic21} and compared to theory predictions.}
\label{fig:4}
\end{figure}
The charmonium system was one of the first physics topics widely studied with the LHC collisions at $\rm 7~TeV$\cite{bib:CMSJpsi}. With increased statistics, the attention moved to the $\Upsilon$ system, triggered and reconstructed through the clean $\mu^+\mu^-$ signature. ATLAS observed a new state\cite{bib:ATLchib3p}, interpreted as $\chi_b(3P)$, at $\rm M=10.530\pm 0.005 (stat.)\pm 0.009 (syst.)~GeV$ in radiative transitions to $\Upsilon(1S)$  and $\Upsilon(2S)$. 
CMS measured the polarization of $\Upsilon(1S)$, $\Upsilon(2S)$ and $\Upsilon(3S)$ separating the data in two intervals of $\Upsilon$ rapidity and five transverse momentum bins\cite{bib:CMSYpol}. The polarization parameters\cite{bib:polpar} $\lambda_\theta, \lambda_\phi$ and $\lambda_{\theta\phi}$ are extracted from the angular distributions of the polar and azimuthal angles of the $\mu^+$ with respect to the z axis of the chosen polarization frame. Following a recently recommended strategy\cite{bib:polpar}, the polarization parameters are measured in three different polarization frames along with the frame invariant combination  $\lambda=(\lambda_\theta+3\lambda_\phi)/(1-\lambda_\phi)$. For all investigated $\Upsilon$ states no sign of significant transverse or longitudinal polarization is observed, in agreement with the findings of CDF, while the theory predictions are often predicting large and strongly model dependent polarization effects. The comparison between CMS results, CDF measurements and a couple of theoretical predictions from perturbative QCD is shown in Fig.\ref{fig:4} (left) for $\Upsilon(3S)$. 

ATLAS has recently produced a measurement\cite{bib:ATLYxsec} of the differential production cross section of $\Upsilon(nS)$ in the approximation of null polarization. It can be seen that the dependence of the acceptance on the polarization can enhance or suppress the estimate of the production rate by up to about a factor of two. The variation affects mainly the contribution to the total cross section coming from the low $\rm p_T(\Upsilon)$ region, while it induces a minor uncertainty at high $\rm p_T$. The importance of the ATLAS measurements lies on the extension of the $\rm p_T$ range which reaches the unprecedented value of 70~GeV. 
The measurements are in agreement with earlier results, including an early CMS measurement. The comparison with theory, shown in Fig.\ref{fig:3} (center), is complicated by the contribution of radiative decays of higher mass charmonium states, which is generally not included in the calculations. The improved NLO non relativistic QCD prediction\cite{bib:nlostar} is in rather good agreement with data in the intermediate $\rm p_T$ region, although it is affected by large uncertainties, arising from the incompete order of the calculation. The color evaporation model\cite{bib:thCEM}, which naturally accounts for phenomena like radiative feed-down, doesn't reproduce well the very low and high $\rm p_T$ regimes. 

Back to the $c\bar{c}$ bound state, CMS has measured the ratio of production cross section times branching fractions for the radiative decay of the $\chi_{c2}$ and $\chi_{c1}$ states to $J/\psi$\cite{bib:CMSchic21}. 
The radiative decays have been reconstructed using photon converted to $e^+e^-$ in the material in order to have good resolution at the very low energy of the photons involved in these transitions. 
After correcting for the known branching fractions, the ratio is found to be in agreement with improved NLO non relativistic QCD calculations.

\section{Studies of the $\rm B_s$ system and searches for rare decays}
\label{sec:bsubs}
\begin{figure}[b]
\begin{minipage}{0.3\linewidth}
\centerline{\includegraphics[width=0.9\linewidth]{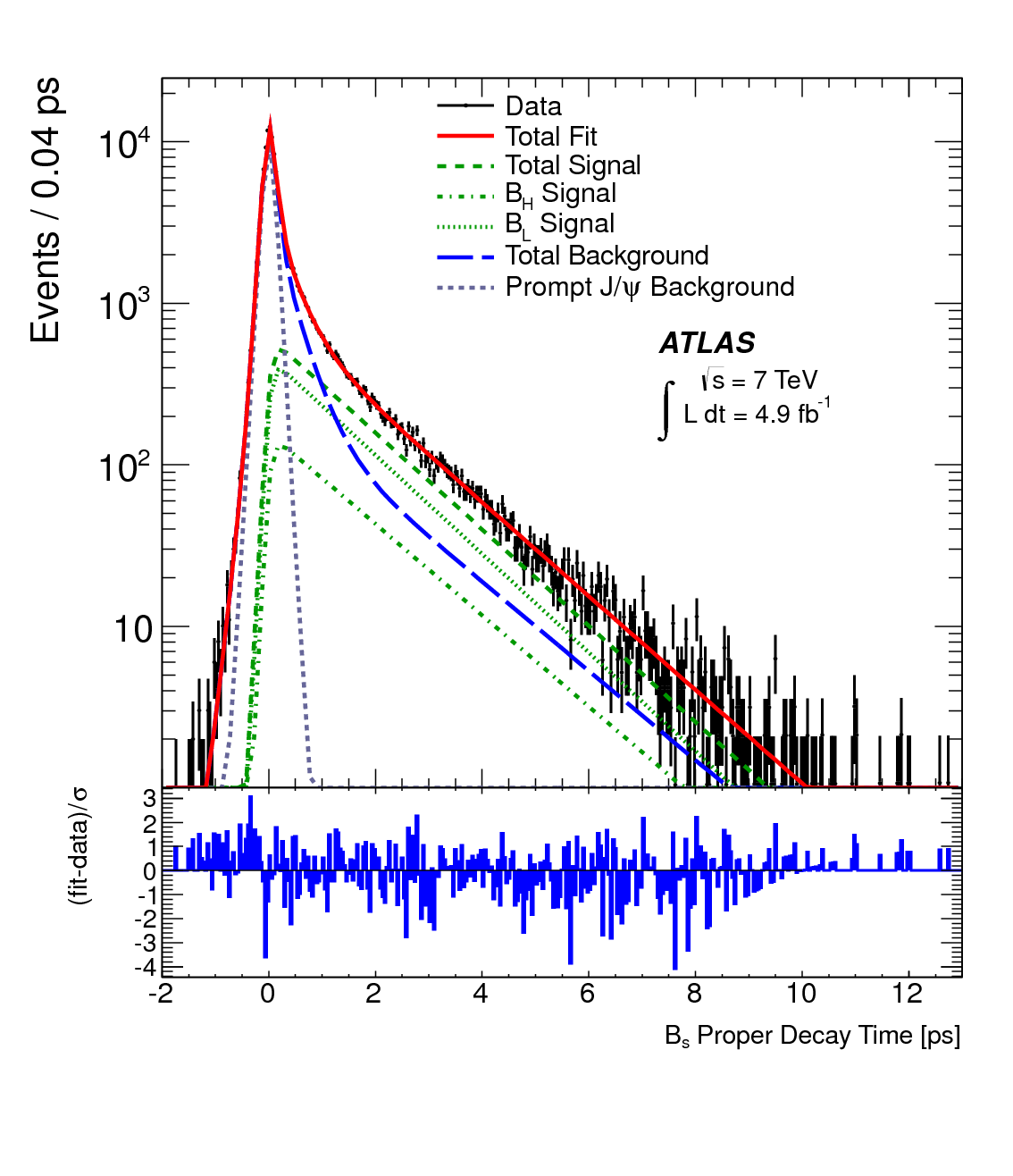}}
\end{minipage}
\begin{minipage}{0.3\linewidth}
\centerline{\includegraphics[width=1.0\linewidth]{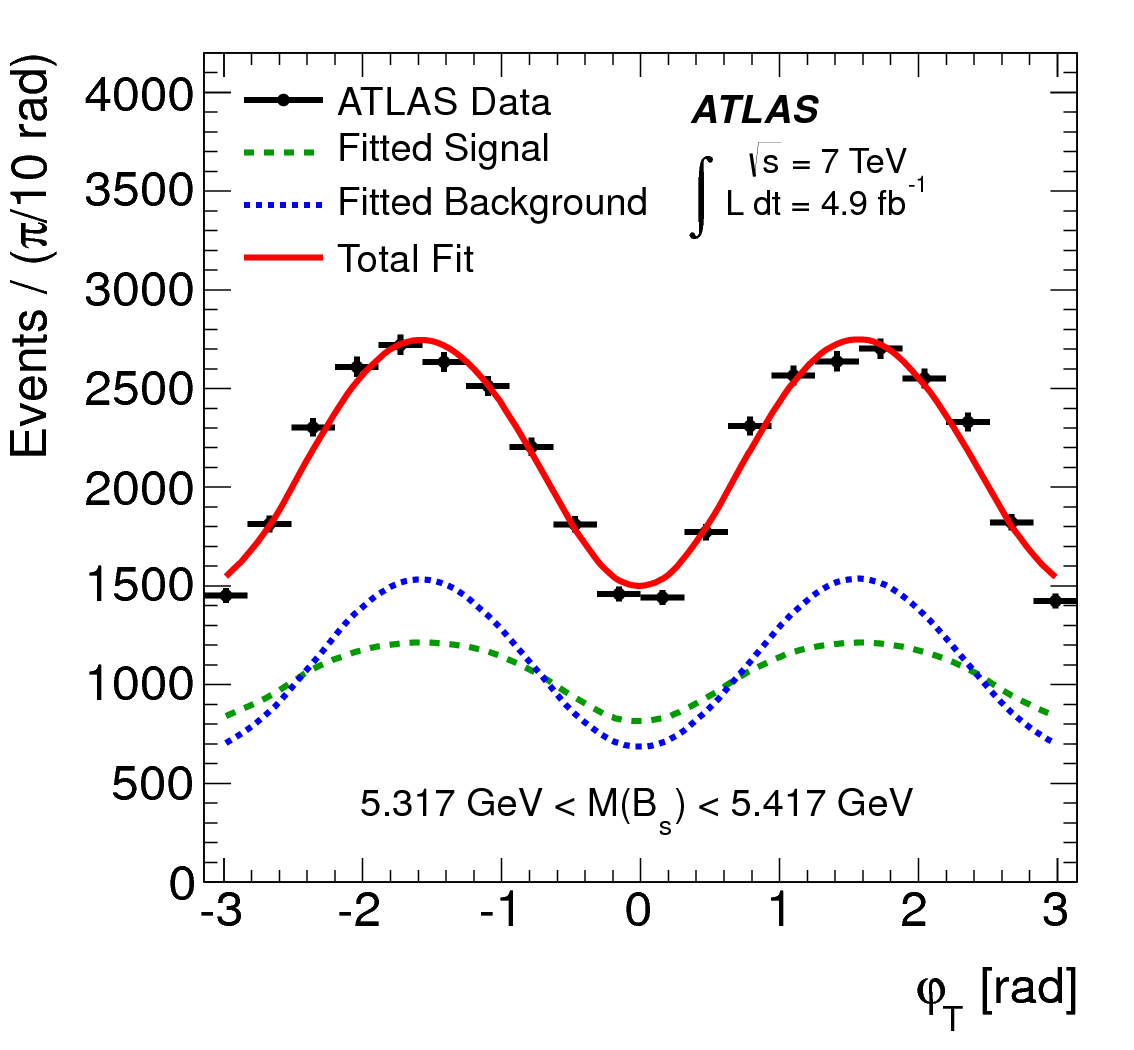}}
\end{minipage}
\begin{minipage}{0.4\linewidth}
\centerline{\includegraphics[width=1.\linewidth]{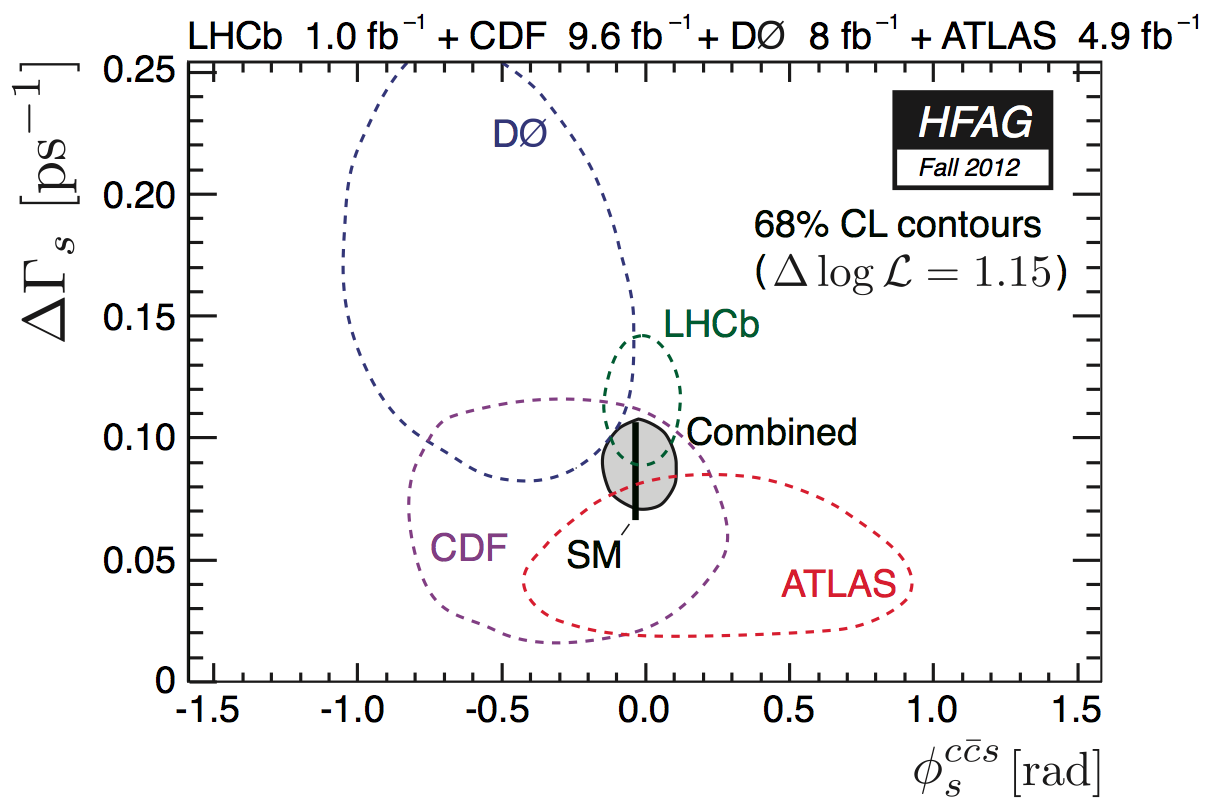}}
\end{minipage}
\vspace{-0.7cm}
\caption{The ATLAS study\protect\cite{bib:ATLbsCP} of the $B^0_s$ system: proper time (left) and $\phi_T$ (center) distributions for the selected candidates; the fit projection is overlaid to the data. The Fall 2012 HFAG combination\protect\cite{bib:HFAG}, including the ATLAS results, of the constraints in the $\Delta\Gamma_s$ and $\phi_s$ plane at 68\% CL (right).}
\label{fig:5}
\end{figure}
$B_s$ can be observed through the fully reconstructed weak decay to $J/\psi \phi$ with $J/\psi\rightarrow \mu^+\mu^-$ and $\phi\rightarrow K^+K^-$. This channel is very important because, being open to both $B^0_s$ and $\bar{B^0_s}$, the interference between decay and oscillation amplitudes gives rise to CP violation via the phase $\phi_s$. In the Standard Model (SM) $\phi_s$ has a small value, $0.0368\pm 0.0018$\cite{bib:phistheory}, directly related to parameters of the CKM matrix, but it is expected to receive extra contributions in several scenarios of new physics.
ATLAS produced an analysis\cite{bib:ATLbsCP} of this decay based on the reconstructed mass, proper decay time ($\tau$) and the three angles, $\theta_T, \phi_T$ and $\psi_T$, defined in the so-called transversity frame, which uniquely define the full decay topology (with no attempt to tag the flavor of the $B^0_s$ meson). 
While the mass allows signal/background discrimination, the proper decay time distribution is sensitive to the lifetime difference between the two mass eigenstates and the angles are sensitive to the CP violating parameter $\phi_s$. The power of this ATLAS measurement lies on the high statistics of the $B_s$ sample extracted from the data. 
The results of the unbinned maximum likelihood fit, projected onto the $\tau$ and the $\phi_T$ distributions in Fig.\ref{fig:5}, 
imply constraints on the $(\phi_s,\Delta\Gamma_s)$ plane which are consistent with those provided by CDF, D0 and LHCb. The Heavy Flavor Averaging Working Group produced a new combination\cite{bib:HFAG}, including the ATLAS constraints in Fall 2012, shown in Fig.\ref{fig:5} (right) where the agreement between the measurements and the SM prediction\cite{bib:phistheory} is clearly established, although more precision is still needed on $\phi_s$. 
A preliminary measurement of the lifetime difference was released by CMS\cite{bib:CMSbs}. The measured value, in agreement with the ATLAS results, is derived under the simplyfing assumption $\phi_s=0$. 

Suppressed flavor changing neutral current decays of hadrons with beauty or charm are suggested as test processes for new physics radiative effects which might enhance the very low rate predicted by the SM. Decays to muon pairs of $B^0_s$ and $B^0_d$ are particularly interesting due to the clean experimental signature, which allows to trigger on such processes in the busy LHC environment. The 2011 data set was used, partially by ATLAS and entirely by CMS, to derive limits on the branching fractions for these decays. The measurement consists on a count experiment in a narrow invariant mass region around the meson mass, after a selection carefully optimized to maximize the single-event-sensitivity. The non-resonant background is estimated through the interpolation of the events in the sidebands, while the resonant $B\rightarrow h^+h^-$ is derived from simulation. Different techniques, typically data driven, are developed by the two experiments to improve the signal sensitivity and the background rejection and, at the same time, to avoid biases from the use of  background data events as a training ground for the selection optimization. The results\cite{bib:CMSbmumu,bib:ATLbmumu}, ${\rm BR}(B_s\rightarrow\mu\mu)<7.7(22)\times10^{-9}$ at 95\% CL  from CMS(ATLAS) and  ${\rm BR}(B^0\rightarrow\mu\mu)<18\times10^{-10}$ from CMS, were combined\cite{bib:LHCbmumu} in the Summer 2012 with the LHCb limits and allowed to refine the more stringent constraints coming from the LHCb data.


%

\end{document}